\documentclass[conference,10pt]{IEEEtran}
\usepackage[dvips]{color}
\usepackage{epsf}
\usepackage{times}
\usepackage{epsfig}
\usepackage{graphicx}

\usepackage{amsmath}
\usepackage{amssymb}
\usepackage{amsxtra}
\usepackage{amsthm}
\usepackage{bbm}
 
\usepackage{amsfonts} 
\usepackage{textcomp}
\usepackage{xcolor} 
\usepackage{multirow}  

\usepackage{here}
\usepackage{rawfonts}
\usepackage{times}
\usepackage{url}
\usepackage{cite}
\usepackage{comment}
\usepackage[utf8]{inputenc}
\usepackage{caption}
\usepackage{subcaption}

\usepackage{pstricks}
\usepackage{algorithm}
\usepackage{algpseudocode}
\usepackage{lipsum}
\usepackage{mathtools}

\usepackage{geometry}
\makeatletter 

\IEEEoverridecommandlockouts

 \def\ps@headings{%
 	\def\@oddhead{\mbox{}\scriptsize\rightmark \hfil \thepage}%
 	\def\@evenhead{\scriptsize\thepage \hfil \leftmark\mbox{}}%
 	\def\@oddfoot{}%
 	\def\@evenfoot{}}
 \makeatother
\begin{document}

\title{ 
	%Application-layer Characterization of Encrypted QUIC Transport Protocol 
Application-layer Characterization and Traffic Analysis for Encrypted QUIC Transport Protocol\\
}
%\vspace{-0.2cm}

\author{ 
	\IEEEauthorblockN{ Qianqian Zhang and Chi-Jiun Su}	
	\IEEEauthorblockA{
		Advanced Development Group, Hughes Network Systems, Germantown, MD, USA\\
		Emails: \url{{Qianqian.Zhang, Chi-Jiun.Su}@hughes.com}. 
		}
}\vspace{-0.2cm}
\maketitle
\vspace{-0.2cm}

\begin{abstract}

Quick UDP Internet Connection (QUIC) is an emerging end-to-end encrypted, transport-layer protocol, which has been increasingly adopted by popular web services to improve communication security and quality of experience (QoE) towards end-users. 
However, this tendency makes the traffic analysis more challenging, given the limited information in the  QUIC packet header and full encryption on the payload.  
To address this challenge, a novel rule-based approach is proposed to estimate the application-level traffic attributes without decrypting QUIC packets. 
Based on the size, timing, and direction information, our proposed algorithm analyzes the associated network traffic to infer the identity of each HTTP request and response pair,  as well as the multiplexing feature in each QUIC connection.   
The inferred HTTP attributes can be used to evaluate the QoE of application-layer services and identify the service categories for traffic classification in the encrypted QUIC connections. %, and diagnose possible network issues in face of congested transmissions.  
\end{abstract}

\section{Introduction}\label{sec_intro}

Passive  monitoring over the network traffic is essential for Internet service providers (ISPs) and network operators to perform a wide range of network operations and management activities \cite{akbari2021look}. 
Given the monitored network status,  ISPs can adjust the capacity planning and resource allocation to ensure a good quality of experience (QoE). 
Network monitoring also facilitates intrusion detection and expedites troubleshooting to guarantee a stable service connectivity for the customers.  
Due to the lack of access to user applications, devices, or servers, passive monitoring is generally challenging. 
As concerns on the privacy violation continually grow, popular applications start to adopt encrypted protocols. 
For example, most prominent web-based services apply hypertext transfer protocol secure (HTTPS)  to protect the security for bi-directional communications between the Internet users and servers. 
Consequently, encryption on the one hand protects users' privacy, but also disables the current network management mechanisms for QoE monitoring and optimization.   

Among all current efforts to incorporate encryption, a new transport-layer protocol, called Quick UDP Internet Connections (QUIC), has emerged to improve communication security and QoE for end-users \cite{langley2017quic}.  
QUIC is a UDP-based, reliable, multiplexed, and fully-encrypted protocol. %, where encryption prevents modification and limits ossification of the protocol over the packet delivery path, and the use of UDP allows QUIC packets to traverse middleboxes. 
As a user-space transport, QUIC can be deployed as part of various applications and enables iterative changes for application updates.  
Compared with Transmission Control Protocol (TCP), QUIC uses a cryptographic handshake that minimizes handshake latency, and  eliminates head-of-line blocking  by using a lightweight data structure called streams,   so that QUIC can multiplex multiple requests/responses over a single connection by providing each with its own stream ID, and therefore loss of a single packet blocks only streams with data in that packet, but not others in the same QUIC connection. 
HTTP-over-QUIC is standardized as HTTP/3 and attracted wide interest from the industry \cite{xu2020csi}. 
Historical trend in \cite{quicUsage} shows that over $7\%$ of websites are already using QUIC, and QUIC is expected to grow in the mobile  networks and satellite communication systems.  

Compared with other encryption technologies, QUIC brings tougher  challenges on passive traffic monitoring.  
For example, TCP header provides useful information, including flags and sequence number, which enable  ISPs to inspect the TCP communication status. 
However, the encryption applied to the QUIC headers leaves very limited information to infer their connection states.  
Meanwhile, %multiplexing and concurrency of HTTP requests and responses over multiple streams in QUIC add complexity to passively evaluate the application performance and user's QoE. 
in the satellite-based network systems,  TCP traffic is usually optimized with Performance Enhancing Proxies (PEPs) \cite{border2020evaluating}. However, QUIC’s end-to-end encryption disables PEP optimizations, which results in an under-performance, compared with TCP PEP, even with QUIC’s fast handshake.  
To address the aforementioned challenges, several recent works in \cite{xu2020csi} and \cite{anderson2019limitless, jain2019application, zhan2021website, tong2018novel, mangla2019using, mazhar2018real, bentaleb2020inferring} have studied the passive monitoring over encrypted  network traffic.  
Authors in \cite{anderson2019limitless} and \cite{jain2019application} investigated the HTTP request and response identification for the application-layer characterization. However, both approaches only support the TCP protocol, which cannot be easily extended to QUIC, due to the limited information in the QUIC transport header. 
Previous works in \cite{zhan2021website} and \cite{tong2018novel} focused on  the QUIC traffic analysis for website fingerprinting and traffic classification. However, both analytic results relied on large-scale statistics of IP packets, but failed to extract the application-layer attributes.  
To infer the application-level information,  the authors in \cite{xu2020csi} and \cite{mangla2019using, mazhar2018real, bentaleb2020inferring} studied the network monitoring for HTTP-based encrypted  traffic, including both TCP and QUIC. Although these works successfully modeled the application-layer QoE for video applications, their approaches cannot be applied to other types of web services, such as web browsing or bulk traffic. %which can have very different patterns from video service.  
Therefore, existing literature shows distinct limitations in terms of QUIC traffic analysis on estimating the application-layer attributes. %, and our work aims to push this boundary by providing a general-purpose framework to understand the encrypted QUIC traffic from the HTTP application level  for various web services.

The main contribution of this work is, thus, a novel rule-based general-purpose framework to explore the application-level traffic attributes without using any decryption towards QUIC header or payloads, for various web services. 
%In particular, based on the size, timing, and direction information visible in the encrypted QUIC packet, our proposed algorithm analyzes the associated network traffic to infer the identity of each HTTP request and response pair and multiplexing feature in each QUIC connection. 
%The proposed approach first infers the packets corresponding to client requests, and then identifies server’s responses, and finally, pairs each request with its associated response, given the network-dependent constraints of inter-packet time and round-trip time (RTT). 
%Once HTTP multiplexing is detected, several requests will be matched as a group with their corresponding responses, to form a super HTTP request-response pair. 
%The proposed algorithm supports both online and offline estimations for HTTP request-response pairs over QUIC protocol, for different applications, including video traffic, web browsing, and bulk traffic. 
%We tested our algorithm under various network conditions, given different maximum transfer size (MTU) and RTT, and the proposed approach gives highly accurate estimation results in both terrestrial and satellite network systems.
%The main contribution of this work is, thus,  a novel rule-based approach to explore the application-level traffic attributes without using any decryption towards QUIC header or payloads. 
Our key contributions include: 
\begin{itemize}
	\item Based on the size, timing, and direction information visible in the encrypted QUIC packet, our proposed algorithm analyzes the associated network traffic to infer the attributes of each HTTP request and response pair, including the start and end time, size, request-response association, and multiplexing feature in each QUIC connection. Once HTTP multiplexing is detected, several requests will be matched as a group with their corresponding responses, to form a super HTTP request-response pair. 
	
	\item The proposed algorithm supports both online and offline estimations for HTTP request-response pairs over QUIC protocol. 
	In the online setting, the real-time traffic is processed by a three-module state machine to determine the instant status of the HTTP request-response communication. 
	In the offline setting,  we consider all QUIC packets at the end of the connection, where 
	the proposed approach first infers the packets corresponding to  client requests, and then identifies server's responses, and finally, pairs each request with its associated response, given the network-dependent constraints of inter-packet time and round-trip time (RTT).  
	
	\item The proposed algorithm can identify QUIC control messages versus HTTP request/response data packets. 	To avoid overestimation of HTTP request/response size, a dynamic threshold on the QUIC packet length is designed to filter out the acknowledgment packets, setting packets and control information in the HTTP traffic when estimating the HTTP request/response data objects.  
	Meanwhile, the proposed algorithm can handle  special features in QUIC protocol, such as  0-RTT request. %	Therefore, the 0-RTT session resumption in QUIC will be identified, and no HTTP request in the 0-RTT packets will be ignored.  

	\item The proposed algorithm  can be applied to different applications, including video  traffic,  website browsing,  interactive web traffic, such as user login authentication, and bulk traffic for file upload and download. 
	We tested our algorithm under various network conditions, given different maximum transfer size (MTU) and RTT, and the proposed approach gives highly accurate estimation results in both terrestrial and satellite network systems.
	%Meanwhile, the proposed algorithm can be applied under various network conditions, given different maximum transfer size (MTU) and RTT. Our method can automatically detect the values of MTU and RTT, and adjust itself for the new setting while estimating the application-layer characterization. We tested our algorithm under various network conditions, and the proposed approach gives highly accurate estimation results in both terrestrial and satellite network systems. 
	
\end{itemize}

The rest of this paper is organized as follows.
Section \ref{sec_system} provides the system overview. The algorithm design ,  including request estimation, response estimation, and request-response match models, is given in Section \ref{sec_algo}. 
In Section \ref{sec_performance}, the performance evaluations are presented, and  Section \ref{discusison} discusses the limitation and future work.  %in Section \ref{sec_app}, we provide possible applications and use cases of our proposed algorithm. 
In the end, Section \ref{sec_conclusion} draws the conclusion. 

\section{System Architecture} \label{sec_system}

\begin{figure}[t]
	\begin{center}  \vspace{-0.2cm}
		\includegraphics[width=8cm]{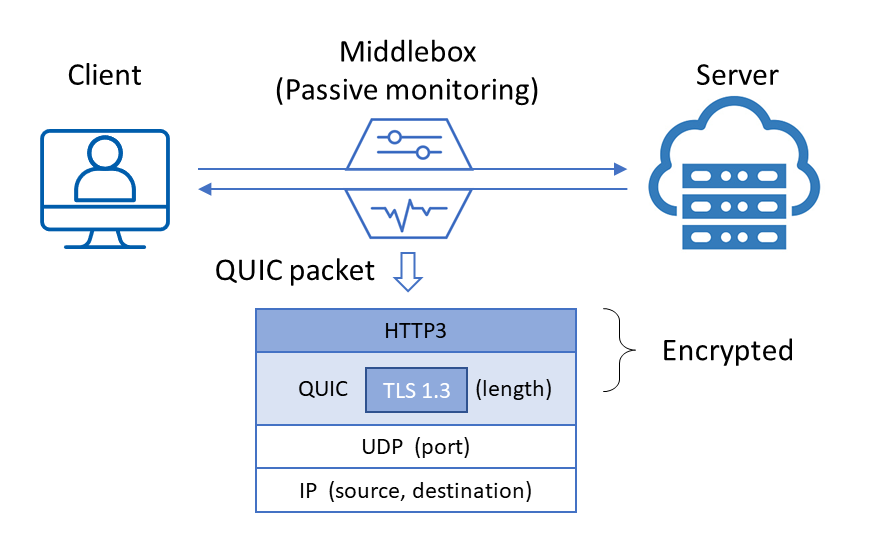} 
		\caption{\label{fig_system}\small Passive monitoring  the bi-directional QUIC packets at a middlebox to infer the application-layer metrics.   }
	\end{center}  \vspace{-0.5cm}
\end{figure}

In this section, we first define the input and output of the traffic monitoring task, and provide a system overview of the QUIC characterization algorithm.  
As shown in Fig. \ref{fig_system}, we consider a passive monitoring module implemented  at a middlebox  between the client and server. 
The middlebox should be able to perceive complete bi-directional traffic without any omission. For example, to observe the traffic of a user, the module can be placed at the user's network access point, while to study the traffic for a cluster of clients, the algorithm can be implemented at the network gateway.  
Relying on the discriminative attributes that is visible in the encrypted QUIC packets, we aim to identify each HTTP pair or HTTP object, consisting of an HTTP request and its corresponding response, which contains key information for the passive monitor to infer the application-layer characterization. % the application-layer metrics for the Internet customers.  

\subsection{Input features} \label{sec_sys_input}

Useful information in the encrypted QUIC packets mainly comes from the network layer and transport layer, including the source and destination  IP addresses, source and destination port numbers,  packet length,  and the limited header information that is still visible in the encrypted packet. 
Meanwhile, packet arrival time and packet position in the sequence of a QUIC flow can also provide essential information for our application-layer characterization. 
In order to support a real-time estimation, the input features require only the information of individual  QUIC packets. 
In our proposed approach, no  window-based feature or large-scale statistical analysis is required. %, such as average data rate over a time window of five seconds, and our proposed approach does not depend on the  either. 
If needed, the window-based feature  can be calculated in the post-processing stage using our estimation results.

\begin{figure*}  
	\centering
	\includegraphics[width=13cm]{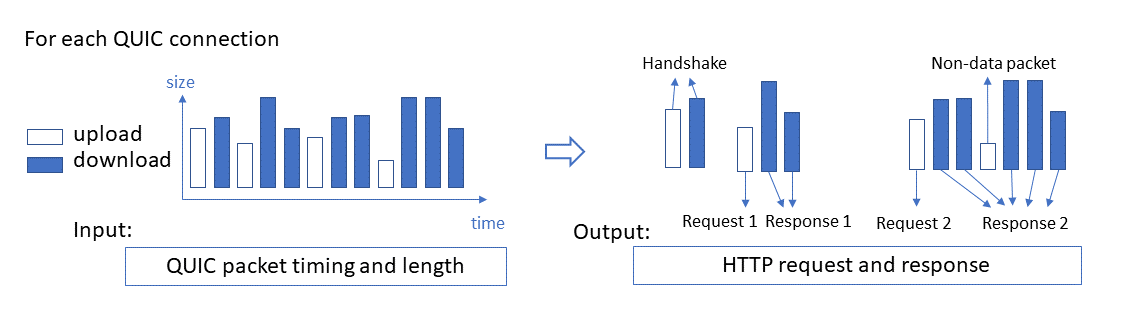} \vspace{-0.15cm}
	\caption{\label{fig_overview}\small  Time and length information of each QUIC packet forms the input to estimate HTTP requests and responses. }  \vspace{-0.25cm}
\end{figure*}

In the network trace, each QUIC connection can be identified by 6-tuples, i.e., source IP, destination IP, source port, destination port, protocol, and QUIC connection ID.  
Within a connection, a sequence of bi-directional QUIC packets with their timing and length information can be observed, and the network operator can extract a small set of features from the network and transport layer headers as the input for the application characterization algorithm,  
which includes \textbf{QUIC header type}, \textbf{QUIC packet length}, \textbf{packet arrival time}, and \textbf{packet order and position},  for upstream and downstream traffic, separately. 
The definition of each input and the reason for choosing these features are given as follows. 

\subsubsection{QUIC header type} \label{sec_sys_input_header}
A QUIC packet has either a long or a short header. 
The most significant bit of a QUIC packet is the Header Form bit, which is set to $1$ for long headers, and $0$ for short headers. 
This header type is always available  in the encrypted QUIC packets and stays invariant across QUIC versions \cite{kuehlewind2020manageability}.  
The long header is used in the handshake stage to expose necessary information for version negotiation and establishment of 1-RTT keys between two ends.  %while the short header is used after connection establishment \cite{kuehlewind2020manageability}.  
Therefore,  the header type provides key information on whether handshake is finished or not.  
Except for 0-RTT resumption, most of the HTTP requests and responses occur only after the handshake is finished. 
Thus, once a QUIC packet with short header is observed in newly-built QUIC connection, soon the HTTP request and response packets are expected to arrive.  

%Therefore, the header type can be used to tell whether handshake is finished or not, and most of the HTTP requests and responses occur only after the handshake stage.
%Thus, once a QUIC packet with short header is observed in newly-built QUIC connection, soon the HTTP request and response packets are expected to arrive.  

\begin{figure} %\vspace{-0.2cm}
	\centering
	\includegraphics[width=7cm]{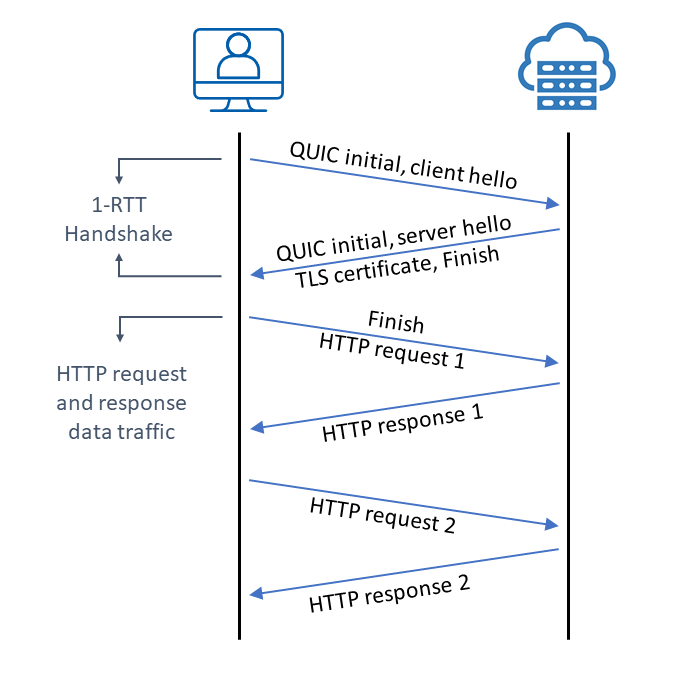} \vspace{-0.15cm}
	\caption{\label{fig_time}\small   1-RTT handshake and HTTP transmission.  } \vspace{-0.4cm}
\end{figure}

\subsubsection{QUIC packet length}  \label{sec_sys_input_len}  
The QUIC packet size can be used to infer whether a QUIC packet contains HTTP data content. %or not. %non-data information.   
First, let us define what an HTTP data packet is. 
Typically, HTTP communications follow a pattern of client's request first, and then server's response. 
Thus, the QUIC protocol for HTTP web applications always uses \textbf{client-Initiated bidirectional stream} \cite{iyengar2020quic}, with a stream ID  that is a multiple of four in decimal, i.e., 0, 4, 8, and etc.  
And the corresponding response will be transmitted over the stream with the same ID as its request. 
In this work, we call the client-Initiated bidirectional stream as data stream, and all the other kinds as non-data stream. 
Although the stream ID can provide accurate information to identify HTTP request and response, this information is encrypted  and invisible at the middlebox.  Therefore, to distinguish the QUIC packet with data content, we must rely on the explicit information, such as the QUIC packet length. 

After the handshake stage, a QUIC packet with HTTP data content usually has a larger length, compared with non-data packets. 
For example, acknowledgment (ACK) is a common type of QUIC frames which  usually has a much shorter size.  
Thus, by setting a proper threshold to the QUIC packet length, it is possible to filter out non-data packets. 
%For example, c
Considering a QUIC packet from the server to the client, if its length is smaller than the threshold $L_{\text{resp}} \in \mathbb{Z}^{+}$, then we consider this packet as non-HTTP-response packet, and exclude it from forming the input features of the estimation algorithm.  
A typical value of response length threshold is $L_{\text{resp}}=35$ bytes. 
%Meanwhile, inside the payload of one UDP packet, there can be one or multiple QUIC packets. Therefore, instead of using UDP packet size as fingerprint,  the length of each QUIC packet provides more fine-gained information for accurate estimations.  
Note that, throughout this paper, the packet length specifically denotes the overall size of a QUIC packet in bytes. 

\subsubsection{Packet arrival time} \label{sec_sys_input_time}
The packet arrival time can be used to tell whether two packets belong to the same HTTP object, and whether a QUIC connection is still active. % of individual packets, we can calculate the time difference of any two packets in the same QUIC connection, and the  inter-arrival time between  packets provides essential guidelines for our analysis in the following three  manners. 
First, an HTTP response  usually consists of multiple packets. 
When two sequential packets are transmitted from the server to the client, the middlebox needs to tell whether they belong to the same response or not.  
Thus, a threshold is applied to their inter-arrival time. For example, if the inter-arrival time of two response packets is less than a threshold $\Delta T_{\text{resp}}$,  then these two packets belong to the same HTTP response;  Otherwise, they are associated with two different responses. 
A typical value for $\Delta T_{\text{respe}}$ is one RTT, and a similar threshold $\Delta T_{\text{req}}$ is applied to consolidate or separate  request packets. %, with a common value $\Delta T_{\text{request}} = 1 $ RTT. 

Second, given a detected request and an estimated response, we need to know whether they are associated HTTP request-response pair. 
Here, we propose a necessary requirement, where the time difference between the first response packet  and the first request packet must be greater than one RTT,  but smaller than $20$ RTTs. 
%An example value for $N_\text{timeout}$ is $N_\text{timeout} = 20$. 
If this requirement is not satisfied, %i.e., the response-request time difference is smaller than one RTT or larger than $N_\text{timeout}$ RTTs, 
then the request and  response are not an associated HTTP pair. Instead, they should belong to different HTTP request-response pairs. 

Furthermore, in a QUIC connection,  %been quiet for very long time, such as 
if there is no packet transmission in any direction for more than $20$ RTTs, then, we consider the QUIC connection as idle. 
In the idle state, before any new request is observed, all response packets from the server to the client will be disgarded. 
%In summary, packet arrival time provides essential information to consolidate individual packets into request-response pairs, match HTTP requests with their corresponding responses, and check whether a QUIC connection is active or not. 

\begin{table}[t]
	\caption{\label{table_input} Input features}\vspace{-0.1cm}
	\renewcommand{\arraystretch}{1.5}
	\begin{center}  
		\begin{tabular}{|p{2.3cm}|p{5.5cm}|}
			\hline
			 \textbf{Input features} & \textbf{Purposes}   \\
            \hline
			Packet direction &  Separate request and response packets. \\
			\hline
			QUIC Header type &  Check whether handshake is finished. \\
			\hline
			QUIC Packet length & Check whether a QUIC packet contains HTTP request or response data.\\
			\hline
			Packet arrival time &  Check whether two packets belong to the same object, whether an HTTP request is associated with a response, and  whether a QUIC connection is still active.  \\
			\hline
			Packet position and order &  Build HTTP request-response pairs from a sequence of individual QUIC packets. \\%, and filter out large-sized non-data packets. \\ %Build HTTP request and response objects. \\ %
			\hline
		\end{tabular}  \vspace{-0.3cm}
	\end{center}  
\end{table}

\subsubsection{Packet order and position}   \label{sec_sys_input_position}
The packets' positions in the sequence of QUIC flow can provide guidelines for a middlebox to form HTTP request-response pairs, as shown in Fig. \ref{fig_overview}.   
For example, an HTTP response  usually consists of multiple QUIC packets with  a noticeable pattern. % QUIC traffic displays in both HTTP response and request transmissions.  
Based on our observation, the first response packet usually has a length that is slightly smaller than MTU size; then, the response is followed with a sequence of multiple MTU-sized packets; finally, the response ends with a packet with much smaller size. 
The cause for this special pattern is that the response data content can have a much larger size than one MTU's payload, therefore the content will be separated into multiple data frames and transmitted via multiple QUIC packets. The slightly small length of the first response packet is caused by the combination of control frame and data frame into one UDP payload, while the last packet contains the left-over response content in this transmission which is usually much less than one MTU. %the much smaller size
Note that, this pattern is an empirical summary based on our observation and experience, which may not be always true. Later, we will apply this rule as a baseline to design the estimation algorithm with further details to cope with exceptions, such as the first response packet has a MTU length, or the last packet has a larger size. 
%Therefore, when a middlebox  observes a sequence of QUIC packets with special length and noticable order values, it can separate these packets into several groups, each of which forms an HTTP request or response.  
%Given this observation, 
Therefore, based on the pattern, we can consolidate responses from a sequence of individual response packets, together with the requirement of inter-arrival time threshold, to form a HTTP response object. %mentioned in Section \ref{sec_sys_input_time}. 
A similar pattern can be observed in the HTTP request  as well.  However, since most of HTTP requests have very limited content, whose size is smaller than MTU, thus, most of the HTTP requests consist of a single packet with length smaller than MTU but greater than the request length threshold $L_{\text{req}} \in \mathbb{Z}^{+}$. 
%Moreover, the position information can reveal whether a QUIC packet with a large size contains a data frame or not. 

%Meanwhile, the position information helps the middlebox distinguish non-data packets, even though the non-date packet may have a long length just like a  data frame.  Note that, the type of a QUIC frame is encrypted in the QUIC packet payload, thus not visible  to the middlebox.  
%For example, cryptographic (CRYPTO) frame, used to exchange keys for encrypting in-order stream of bytes, usually yields a long length, due to large cryptographic content.  Thus, a QUIC packet with CRYPTO frame usually has similar size as data-stream frame. In order to distinguish the CRYPTO and data packets, we must rely on the position information of the considered packet. 
%Given that CRYPTO packet only exists during or right after handshake stage, thus we can distinguish CRYPTO packets from HTTP response data, since response packets only appear after at least one request which can be explicitly later than handshake. 
%A similar rule applies to the Handshake Done and New Token (NT) frames, which indicate the end of the handshake but their packet has a short header and a large size, thus, these packets might  be easily mis-classified as an HTTP response if we ignored their position information.   

Till now, we have introduced  four types of inputs, and the rational for choosing these features is  summarized in Table \ref{table_input}. %. Key information of  input features % this section list %, together with the packet direction information, and their underlying insights

\subsection{Output metrics} \label{sec_sys_output}

\begin{table}[t]
	\caption{Output metrics} \vspace{-0.1cm}
	\renewcommand{\arraystretch}{1.5}
	\begin{center}  
		\begin{tabular}{|p{2.1cm}|p{5.8cm}|}
			\hline 
			\textbf{Output type} & \textbf{Estimated output}   \\ 
			\hline 
			& Request start time \\
			& Request size \\
			& Number of request packets\\
			& Response start time \\
			Object-level& Response end time \\ 
			& Response size\\
			& Number of response  packets  \\ 
			& Number of individual HTTP request-response pairs  \\    
			& Max length of last ten ACK packets\\
			\hline 
			& Connection start time\\
			& Connection duration\\ 
			& Total request size \\
			& Total response size\\ 
			{Connection-level} & Total number of request packets \\
			& Total number of response packets \\
			& Number of  individual HTTP request-response pairs \\
			& Number of  estimated HTTP  objects \\
			& Level of multiplexing\\
			\hline
		\end{tabular} 
	\end{center}   \vspace{-0.4cm}
	\label{tab_output}
\end{table}

Given the input features, we aim to design a rule-based algorithm to estimate the \textbf{object-level} HTTP request-response information,  and the  \textbf{connection-level} QUIC link information, by passively monitoring the encrypted packet sequences. 

\subsubsection{HTTP object level output}
An HTTP pair consists of an HTTP request  and its corresponding HTTP response. 
For the request part, our designed algorithm will output the start time, size, and the number of request packets in the estimated HTTP request. 
Similarly, the response output metrics include the start time, end time,  size, and the number of response packets in the HTTP response. 
The reason to exclude the request end time from the output metric is the fact that most HTTP requests consist of single-packet, thus, the end time  of a request usually coincides its start time.  

\begin{figure}  
	\centering
	\includegraphics[width=6.8cm]{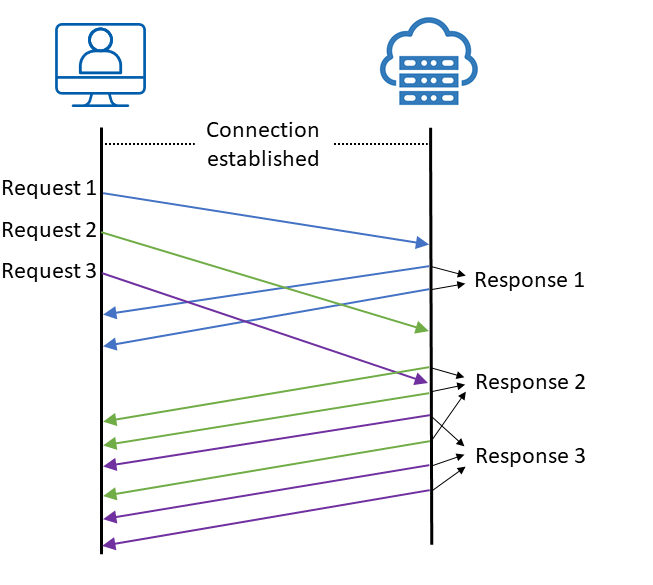} \vspace{-0.25cm}
	\caption{\label{fig_multiplexing}\small  HTTP request-response multiplexing. } \vspace{-0.4cm}
\end{figure}

Since QUIC protocol supports HTTP request and response  multiplexing by creating multiple streams in the same connection, thus, we will see in Fig. \ref{fig_multiplexing} that before the transmission of an existing HTTP response is finished, another request can be sent from the client to server using a new stream ID. 
%s shown  before the client receives response 1, requests 2 and 3 have been sent out. 
In the case of multiplexing, the sequence of request or response packets belonging to different HTTP objects may be interleaved with each other, thus, it might be impossible to separate the packets for each individual HTTP object, based on their length and timing information only.  
%Different from Fig. \ref{fig_time} where multiple HTTP objects can be clearly separated in the timeline,  multiplexing in Fig. \ref{fig_multiplexing} results in a more complicated pattern. 
%For example, the transmission of HTTP response 2 and response 3 are mixed together. 
%Also, if without decryption, it is hard to tell  from the timeline when response 1 is finished.  
%As a consequence, to deal with HTTP multiplexing in a QUIC connection, 
In this case, we will group the interleaved HTTP request-response objects together to form a super HTTP object. 
%In this case, 
And, the output meaning of an estimated super object changes slightly, where 
the request (or response) start time is the time stamp of the first request (or response) packet in the super object, 
the response end time is the time stamp of the last response packet, 
the request (or response)  size is the total size of all request  (or response) packets in the super object, 
and the request (or response) packet number is also the total number of all request (or response) packets. % in the super object.  
Moreover, the number of HTTP pairs denotes the number of individual HTTP request-response pairs grouped in the super object, and only in the case of multiplexing, this value is greater than one. %equals to the number of detected individual requests, and if the current object has no multiplexing, then this value is set to be one. 
When HTTP multiplexing happens, the response estimation can be very confusing, but the request detection is still reliable, thus the number of detected requests is counted to represent the number of individual HTTP pairs in the super object. 

Lastly, the length of the ACK packets contains meaningful information for packet filtering. 
If a packet loss is detected at the client side and the lost packet contains key information that requires re-transmission, then the client will inform the server with the loss information by sending an ACK packet. 
If the number of lost packets keeps increasing,  the ACK frame needs to contain more information, which yields an increased packet length. 
Therefore, by monitoring the ACK packet length in a real-time manner,  the passive observer can accurately determine the threshold for the HTTP data packets, and filter out the non-date frames properly.  
Usually, we keep the length information of the last ten ACK packets for both directions. 
%, based on the sizes of the last few ACK packets. 
%Second, the ACK length indirectly indicates  the level of packet loss or traffic congestion happening in the uplink and downlink. % in terms of application data transmission. 

%In summary, the object-level output for each HTTP (super) request-response pair includes: (a) the request start time, size, number of request packets; (b) the response start time, end time, size,  number of response packets; (c) the number of HTTP pairs, and the maximum length of the last ten ACK packets in each direction. 

\subsubsection{QUIC connection level output}

Once a QUIC connection has been quiet for more than $20$ RTTs, we consider it as inactive, and the overall HTTP transmission will be summarized into a QUIC-connection output, and after that, all memory for this connection will be cleared.    
The connection-level output is shown in Table \ref{tab_output}, where
the connection start time is the timestamp of the first packet from client to server, 
the connection duration is the time different between the first packet to the last over the QUIC connection, 
the total request (or response) size is the length sum of all HTTP request (or response) data packets,  
the total number of request (or response) packets counts the number of all HTTP request (or response) packets in the QUIC connection,  
the number of individual HTTP pairs equals to the number of detected requests, 
and the number of  estimated HTTP objects equals to the number of object-level outputs estimated within this QUIC connection.  
For example, in Fig. \ref{fig_multiplexing},  the number of individual HTTP  pairs is three, while  the number of estimated HTTP objects is only one, due to multiplexing, and in Fig. \ref{fig_time}, the number of individual HTTP pairs and the number of estimated objects both equal to two. 
In the end, we define the level of multiplexing as the ratio of the number of individual HTTP pairs to the number of estimated HTTP objects. 
The value of the multiplexing level ranges in  $[1,  N_\text{req}]$, where $N_\text{req} \in \mathbb{Z}^{+}$ denotes the maximum number of individual HTTP pairs that our algorithm can tolerant in each super object. 
When multiplexing happens, the level of multiplexing is greater than one; otherwise, its value equals to one. 
Here, the level of multiplexing helps 
%contains essential information for 
a network operator to classify the traffic category of a QUIC connection. For example, a web-browsing link usually has a higher multiplexing level than a video link. 
Key information of object-level and connection-level outputs is  summarized in Table \ref{tab_output}.

\section{Algorithm and Approaches} \label{sec_algo}
In this section, we aim to design a rule-based algorithm so that given the input features in Table \ref{table_input}, we can estimate the output metrics in Table \ref{tab_output}. 
To this end, we design a state machine with three modules, 
where the request estimation module infers the client requests, the response estimation module consolidates the QUIC packets into server's responses,  
and  a match module pairs the estimated requests with their corresponding responses, under the network-dependent constraints of inter-arrival time and RTT.  
% Meanwhile, special features in QUIC protocols, including 0-RTT request and HTTP multiplexing, will be handled properly in our designed modules. 
Furthermore, to extend the application range and increase the robustness of our algorithm, three supporting modules are introduced to automatically adjust the threshold for data packet size, detect the MTU size, and estimate the value of RTT, respectively, so that the proposed algorithm supports an accurate estimation in various network systems under different communication conditions.  

\subsection{Request estimation} \label{sec_algo_req}

\begin{figure}  
	\centering
	\includegraphics[width=7cm]{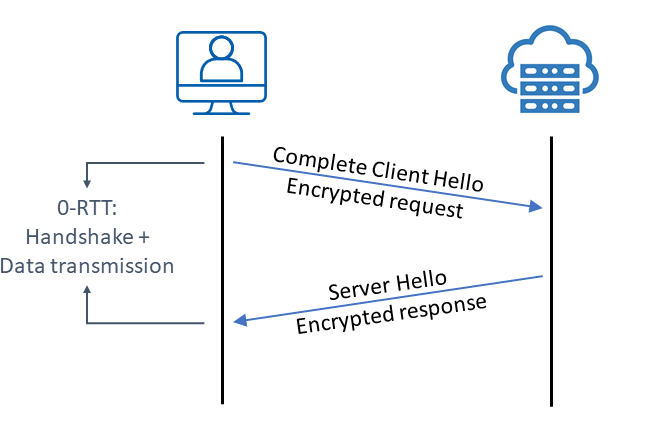} \vspace{-0.25cm}
	\caption{\label{fig_0rtt}\small  0-RTT connection resumption. }  \vspace{-0.25cm}
\end{figure}

In the QUIC protocol, a special features, called 0-RTT connection resumption, is shown in Fig. \ref{fig_0rtt}. 
Assume a client and a server had previously established a QUIC connection, then when a new connection is needed,  client can send application data with the first packet of Client Hello and reuse the cached cryptographic key from  previous communications.  
Notably this allows the client to compute the private encryption keys required to protect application data before talking to the server,  thus successfully reduces the latency  incurred in establishing a new connection. 
%The updated cryptographic key for the new QUIC connection can be exchanged between the client and server in the following communication stage without interrupting the HTTP request and response transmission. 
%, along with the first packet of Initial Hello, a request can be transmitted immediately without waiting for any response from the server. 
Thus, in the case of 0-RTT resumption, the HTTP request and response can happen before the handshake is finished, and a special detection mechanism is needed to infer the 0-RTT request packets. 
%Meanwhile, in this section, we will also show how to detect the HTTP request object after the handshake is finished. 
Given a QUIC packet with a long header,  the third and fourth significant bits in the header indicate the type of this packet. 
%This type field is unencrypted information and stays available to the middle box for the current QUIC version-1  protocol  \cite{iyengar2020quic}. %tells its packet type. 
If the type field shows (0x01), then the packet is a 0-RTT packet \cite{iyengar2020quic}. 
Next, to determine whether a 0-RTT packet contains HTTP request data, we propose three criteria: %, which, based on our observations, hold for most of 0-RTT request packets and vice versa. 
First, a 0-RTT request has usually a single packet;  Second, the  length of a 0-RTT request packet often ranges within $[100,1000]$; Third, there is only one QUIC packet in the UDP payload. 
If all above requirements are satisfied, we can say that this 0-RTT packet is  a 0-RTT request, otherwise,   
%n the contrary, if the length of the QUIC packet is smaller than $100$ or larger than $1000$, or there are more than one QUIC packet in the UDP payload,  then
 this 0-RTT packet is more likely to contain control information, other than HTTP request data. 
Again, these criteria are empirical, which may not always be true. However, according to our observation and experience, the criteria lead to a high accuracy to estimate 0-RTT requests.

Once handshake is finished,  QUIC packets will start to use short headers, which do not have a packet-type field anymore. 
But, similar to 0-RTT requests, request after handshake requires only one QUIC packet in the UDP payload.  
Meanwhile, %to determine whether a packet from client to server contains HTTP request data or not, we propose two criteria, which, based on our observations, hold for most of QUIC request packets and vice versa. 
%First, there is only one QUIC packet in the UDP payload;   
%If all above requirements are satisfied, we can say that this 0-RTT packet is  a 0-RTT request.  
%On the contrary, if the length of the QUIC packet is smaller than $100$ or larger than $1000$, or there are more than one QUIC packet in the UDP payload, then this 0-RTT packet is more likely to contain control information, other than HTTP request data.
%Second, 
the length of a request packet ranges between $L_\text{req}$ and $L_\text{MTU}$, where $L_\text{req}$ is the length threshold for request packets, and $L_\text{MTU}$ is the size of MTU.  
%In particular, the MTU value $L_\text{MTU}$ is usually network and device dependent. Thus,  a supporting module is given in  Section \ref{sec_algo_support_MTU} to automatically detect $L_\text{MTU}$ for each QUIC connection in each communication direction.  
In general, the MTU value $L_{\text{MTU}}$ is network and device dependent with a value range of  $[1200,1360]$. 
Meanwhile, the value of $L_\text{req}$ is  dynamic over time. % within a single QUIC connection in a single communication direction.   
% For example, w
When we are inferring for the first packet of the first request,  the request size threshold is set as $L_\text{req} = 100$ bytes. 
Then once the  first request packet has been detected, the value of $L_\text{req}$ is adjusted to $50$ bytes. 
Later, as the HTTP request transmission continues, $L_\text{req}$ will be dynamically adjusted based on the real-time traffic conditions. 
% to deal with the potential congestion scenario in the communication system. 
Details for adjusting  $L_\text{req}$ will be shown in Section  \ref{sec_algo_thre}. 

\begin{comment}

In summary, after handshake, if the middlebox observes a QUIC packet from the client to the server, having only one QUIC in the UDP payload, with a length between $L_\text{request}$ and $L_{\text{MTU}}$, then this packet will be classified as a request packet. 
 
\begin{figure}[t]
	\begin{center}  
	\includegraphics[width=7cm]{request.png}
	\caption{\label{fig_request} An example of state machine for  HTTP request estimations, where  -1 is initial state, 0 is idle state, 0.5 is waiting state, and 1 is transmission state. 
    Once the algorithm comes to state -0.5 or state 0, a request  is estimated, and the estimated request will be given to the match module to form HTTP request-response pairs.   }
\end{center} \vspace{-0.5 cm}
\end{figure}
\begin{figure}[t]
	\begin{center}  
		\includegraphics[width=7.38cm]{response.png}
		\caption{\label{fig_response} }
	\end{center} \vspace{-0.5 cm}
\end{figure}

\begin{figure}[t]
	\begin{center}  
		\includegraphics[width=8.5cm]{match.png}
		\caption{\label{fig_match} An example of state machine for  HTTP request-response match module, where -1 is initial state, 0 is idle state, 1 is waiting-for-response state, 2 is waiting-for-new-request state, and 3 is waiting-to-output state.  Once the algorithm moves over a double-line arrow, an HTTP request-response pair is estimated, where the  dash-line arrow requires certain conditions for an HTTP output, while the solid-line will give an output without condition. }
	\end{center} \vspace{-0.5 cm}
\end{figure}

\end{comment}

\begin{figure}  
	\centering
	\includegraphics[width=7cm]{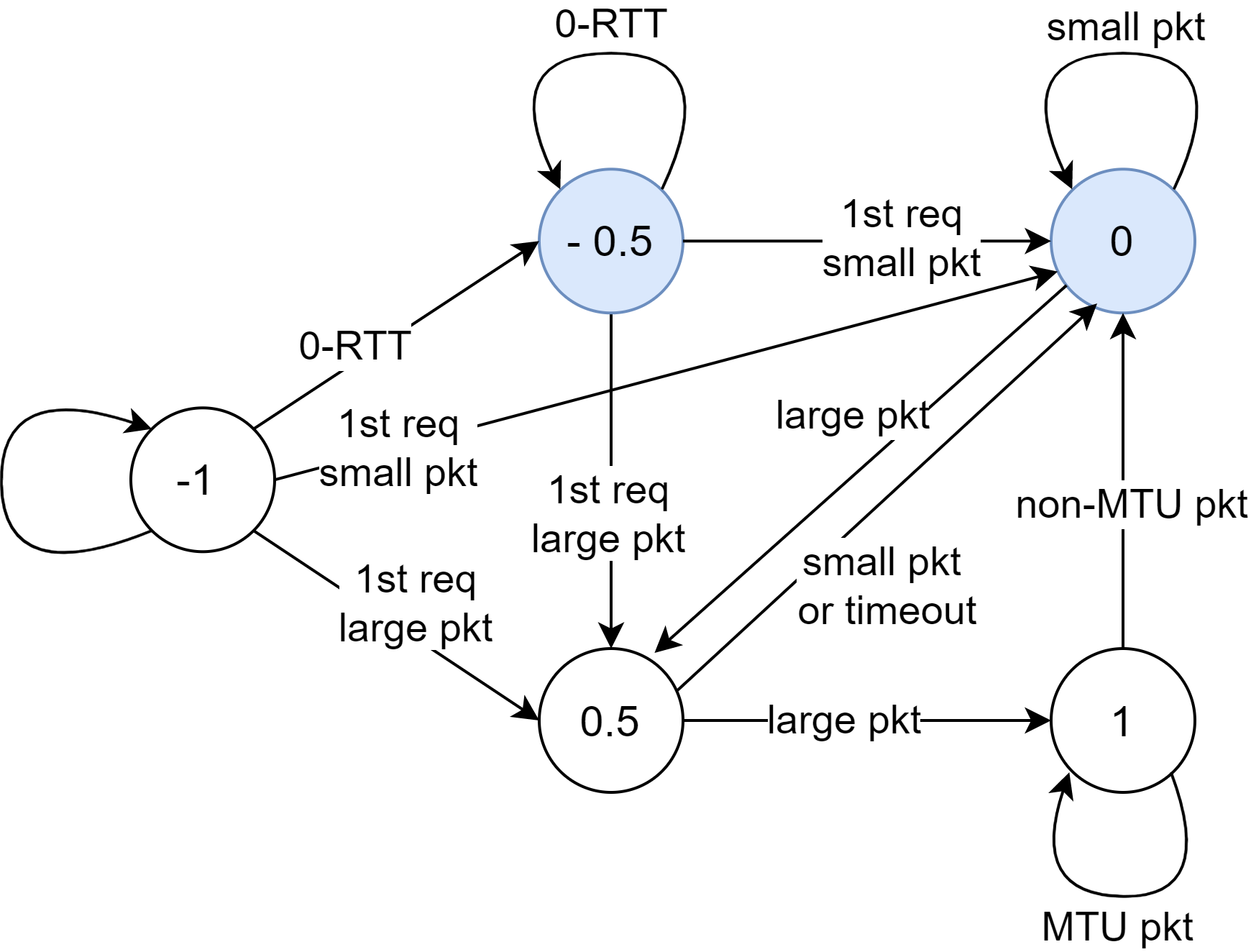} %\vspace{-0.5cm}
	\caption{\label{fig_request}\small   State machine for request estimations, where  -1 is initial state, 0 is idle state, 0.5 is waiting state, and 1 is transmission state. Once the algorithm comes to state -0.5 or state 0, a request is estimated and will be given to the match module.  }\vspace{-0.3cm} % 
\end{figure}

Given that an HTTP request consists of either a single packet or multiple packets, in order to consolidate a sequence of  request packets into a group of request objects, we  design a request estimation algorithm with a state machine shown in Fig. \ref{fig_request}. 
%In order to disclose  the communication stage, in this example, we use the negative value for a state if the handshake is not finished at this moment, while positive values are used for after-handshake states. 
When the client sends the first Initial Hello packet to the server, a state machine is initialized for the QUIC connection with an initial state $-1$. 
During the handshake stage, if a 0-RTT request is detected, the algorithm goes to state $-0.5$. 
As long as the algorithm comes to state $-0.5$,  a 0-RTT request  will be output, and the estimated 0-RTT request  will  be given to the match module. %, detailed in Section \ref{sec_algo_match}, to form HTTP request-response pairs. 
On the other hand, if no 0-RTT request is found, the state will stay at $-1$, until handshake is finished and a new request packet is detected.  
If the new request packet has a length greater than $L_{\text{MTU}} -  8$,  then we consider its as a large packet, and the algorithm will move to state $0.5$. 
Otherwise, if the packet's length ranges in $[L_{\text{req}}, L_{\text{MTU}} - 8]$, we consider it a small request packet, and the algorithm comes to state $0$. 
%Usually, the value of $\Delta L$ is a small and positive integer, such as  $\Delta L = 8$. % in the aforementioned example. 

State $0.5$ is a waiting state, where we need the information of the next request packet to determine whether we are estimating a single-packet  or multi-packet request. 
Therefore, at state $0.5$, if we receive a large packet within one RTT, then, the current request is a multi-packet request, and more packets belonging to the same request might arrive soon, thus, the algorithm moves to the transmission state $1$; 
otherwise, if we receive another small packet at state $0.5$, the estimated request consists of two packets, and the algorithm goes to state $0$, and outputs the estimated request. 
%Note that, as long as the request estimation module moves from state $0.5$ to state $0$, the algorithm outputs a 
%Then, this request will be given the match module to form HTTP request-response pairs.
Meanwhile, if no new  packet arrives within one RTT, then it is a single-packet request. Thus, the algorithm moves to state $0$, and outputs the single-packet request.  
State $0$ is an idle state, meaning  no on-going transmission at this stage. %As long as the module comes from any other state to state $0$,  an estimated request will be output. 
%Staying at state $0$ indicates that all existing requests have been estimated and given to the match module to form HTTP request-response pairs, and the algorithm is waiting for new packets to arrive so that  the new estimation can start. 
At state $0$, if a large request packet comes, the algorithm moves to state $0.5$ to wait for more packets. 
Otherwise, the algorithm will output a single-packet request, and stays at state $0$. % consisting of one single small packet.   
Lastly, state $1$ is a transmission state, meaning a multi-packet request is transmitting a sequence of MTU-sized packets. %This is the second stage of the packet sequence pattern previously mentioned in Section \ref{sec_sys_input_position}.  
At state $1$, if the arrived packet has a MTU-size, then transmission is on-going and the algorithm stays at state $1$. 
If the new packet has a length smaller than MTU, then the transmission of the current request is done, so the algorithm moves to state $0$, and outputs the estimated multi-packet request. %, and gives it to the match module. 

In summary, the request estimation module monitors all QUIC packets from client to server, processes the header, time, length, and order information of each  packet, and outputs the estimated request to the match module. %An example of flow chart for the request estimation algorithm is given in Fig. \ref{fig_reqFlow}. 

\subsection{Response estimation}\label{sec_algo_resp}

\begin{figure}  
	\centering
	\includegraphics[width=7cm]{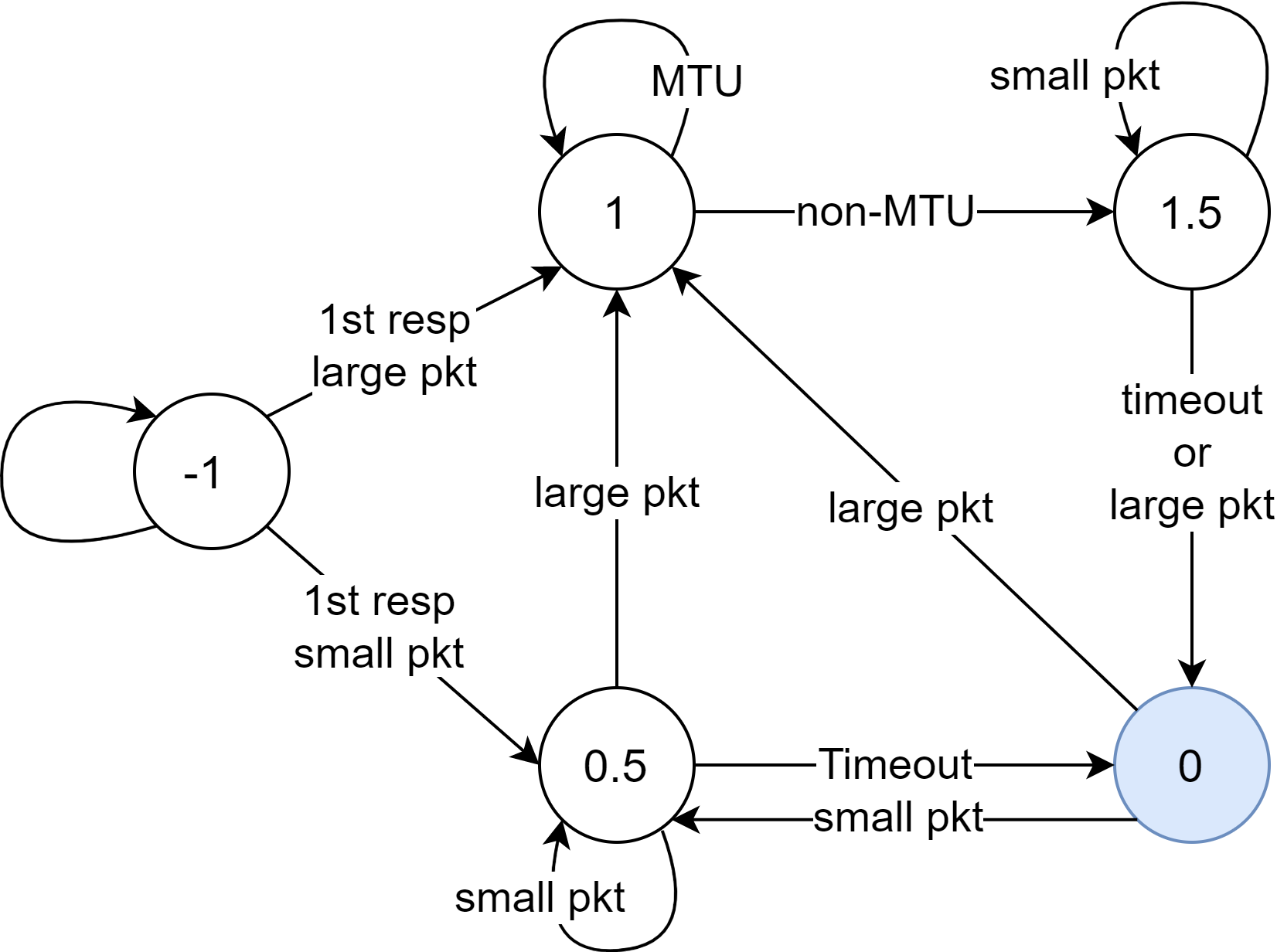} %\vspace{-0.5cm}
	\caption{\label{fig_response}\small  State machine for  response estimation, where  -1 is initial state, 0 is idle state, 0.5 is waiting-to-start state,  1 is transmission state, and 1.5 is waiting-to-end state.  Once the algorithm comes to state 0, a response is estimated and will be given to the match module.   } \vspace{-0.2cm}
\end{figure}

%Given a sequence of QUIC packets from a server to the client over a considered connection,  first we will distinguish whether a packet contains HTTP response data in Section \ref{sec_algo_resp_pkt}, and then, Section \ref{sec_algo_resp_statemachine} will show how to consolidate a sequence of response packets into a group of HTTP responses. 
%\subsubsection{Inferring HTTP response packet} \label{sec_algo_resp_pkt}

Similar to the request packet, an HTTP response packet usually has only one QUIC packet in the UDP payload, and the response packet length ranges between $[L_\text{resp},L_{\text{MTU}}]$, 
%Note that, the MTU size for request and response on the same QUIC connection can be different, but in most cases, they are of the same value, so we will not specify the request-MTU and the response-MTU in this discussion. 
where $L_\text{resp}$ is a dynamic threshold with the initial value of $L_\text{resp} = 35$, and the updating rule will be shown in Section \ref{sec_algo_thre}. 
%Therefore, after the middlebox has detected at least one request,  then if a QUIC packet from the server to the client has only one QUIC in the UDP payload, with a length between $L_\text{response}$ and $L_{\text{MTU}}$, then this packet will be classified as an HTTP response packet. 
%\subsubsection{State machine for response estimation module}\label{sec_algo_resp_statemachine}
%Similar to request module, in order 
To consolidate individual packets into HTTP responses,  a response estimation algorithm is designed in Fig. \ref{fig_response}.  
%However, different from request, most of HTTP responses consist of more than one packet. Therefore, in the response estimation, we need to detect the start and the end of each multi-packet response  more carefully.  As a result, two waiting states are introduced in the response state machine, which are waiting-to-start and waiting-to-end states, respectively. 
Initially, when  no request is detected, the response module stays at state $-1$.  
When at least one request  and a new response packet are detected, the algorithm moves to state $1$ if the response packet size is larger than $L_{\text{MTU}}-8$, or the algorithm moves to state $0.5$ if  the  packet length between $[L_\text{resp}, L_{\text{MTU}}-8]$.

State $0.5$  is a wait-to-start state, meaning after receiving a small packet, we need to see the next  packet to determine  whether it is a single-packet or multi-packet response. % or it is the start of a multi-packet response, thus,  the transmission pattern. 
Therefore, at state $0.5$, if a large packet arrives within one RTT,  the algorithm will move to state $1$; 
if a small response packet arrives within one RTT, the algorithm stays at state $0.5$, and  groups the received small packets into one object. 
Due to different  implementations, some servers may start the multi-packet response with more than one non-MTU packets.    
%On the other hand, if the new packet received at state $0.5$ is a small packet, i.e., QUIC packet length within $[L_\text{response}, L_{\text{MTU}} - \Delta L)$, then, we  group the small packets received before and at state $0.5$ together into one response. 
%There are two reasons for this  small-packets grouping design. 
%First, we observed in many scenarios that a multi-packet response can start with more than one small packet, and then, followed by a sequence of MTU-sized packets; Second, we found that the complex  pattern of response transmissions usually leads to a larger number of responses in our estimation than in the ground truth.  In order to cope with the observed traffic pattern and estimation results, we tend to group several small sequential response packets together into one response. 
If no packet arrives during one RTT,  the algorithm  moves to state $0$, and output an estimated response. % consisting of one or several small response packets. %, received before and at state $0.5$. 
State $0$ is an idle state, meaning no transmitting response. 
%As long as the  algorithm comes to state $0$,  an estimated response  will be output, and this response  will be given to the match module to form HTTP request-response pairs. 
%Staying at state $0$ indicates that all exiting responses have been detected and passed to the match module, and the algorithm is waiting for new packets to arrive and start new estimations. 
At state $0$, if a large response packet comes, the algorithm moves to state $1$. Otherwise, the algorithm comes to state $0.5$. 
State $1$ is a transmission state,  meaning a multi-packet response is transmitting a sequence of MTU-sized packets. At state $1$, if the arrival packet has a MTU-size, then the response transmission continues and the module stays at state $1$.  
Otherwise, the transmission finishes  and  the algorithm moves to state $1.5$.

Lastly, state $1.5$ is a wait-to-end state. 
%This state is designed to cope with the situation where a response can end with more than one small packet. 
%As mentioned in  Section \ref{sec_sys_input_position}, a standard packet sequence of a response should  end with one packet with a length smaller than MTU.  %start with one packet with length slightly smaller than MTU, then followed by a sequence of MTU packets, and then,
%However, if some response packets are missing during their transmission and re-transmission of these packets happens, then 
Due to re-transmission, an HTTP response can end with  multiple small packets. 
%Unfortunately, in the QUIC protocol, the re-transmission packet cannot be easily detected without decryption, since there is no visible field in the QUIC headers to distinguish re-transmission packets from normal response packets.  
%Therefore, in our analysis, we include all response packets having the same stream ID to form the  ground-truth response. 
%Consequently, 
Therefore, at state $1$, if the middlebox observes a small packet, % with length between $[L_\text{response}, L_{\text{MTU}} - \Delta L)$, it is hard to tell whether the response transmission is finished, or there will be more small packets to come. 
%Therefore, the algorithm 
it waits at state $1.5$ for one RTT, during which if more small packets arrive, the algorithm consolidates these small packets with the previous MTU sequence to form one response, and  stays at the state $1.5$ until one-RTT timeout. 
If no packet arrives within one RTT, the response transmission is finished, and the module moves to state $0$ to output the estimated response. 
However, if a large packet arrives within one RTT, then we realize that the previous response has finished, and a large packet belonging to a new response  is received. In this case, the algorithm  first moves to state $0$, output a response  consisting of all packets but not including the last one, then,  the newly-arrived large packet will  start another response  estimation, and move the algorithm to state $1$.  

Thus, the response estimation module monitors all QUIC packets from server to client, processes the header, time, length, and order information of each encrypted packet, and outputs the estimated response  to the match module. % An example of flow chart for the response estimation algorithm is given in Fig. \ref{fig_respFlow}. 

\subsection{Request-response matching} \label{sec_algo_match}

\begin{figure}  
	\centering
	\includegraphics[width=3.8cm]{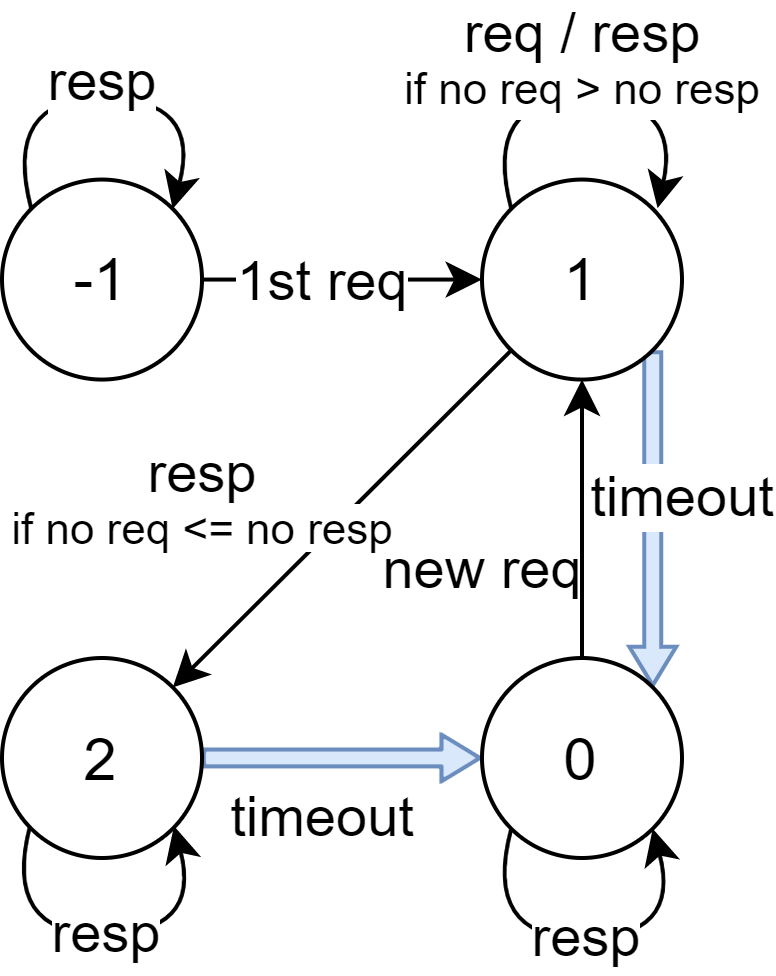} \vspace{-0.05cm}
	\caption{\label{fig_match}\small   State machine for match module, where -1 is initial state, 0 is idle state, 1 is waiting-for-response state, and 2 is waiting-to-output state.  Once the algorithm moves over a double-line arrow, an HTTP request-response pair is estimated. } \vspace{-0.5cm}
\end{figure}

Given the estimated requests and responses, the final step is to match each request and its corresponding response to form an HTTP pair, a.k.a. HTTP object.  
%As mentioned in Section \ref{sec_sys_output_obj}, the QUIC protocol supports HTTP multiplexing, where within one QUIC connection, multiple requests or responses can be transmitted at similar timestamps over different streams. 
%The consequence of multiplexing is that a passive observer cannot separate different individual HTTP objects  using only time and length information. 
%As a result,  we  have to group multiple requests and responses together to form a super HTTP object in case of multiplexing. 
%Fortunately, our request estimation algorithm usually gives a very accurate estimation on the number of requests over the encrypted QUIC connection, therefore, we can clearly know how many individual HTTP objects are there in a super HTTP object, based on the estimated request number.  
%In order to clarify the matching approach, we give an example of the 
A  state machine of the matching module is given in Fig. \ref{fig_match}, which takes the estimated HTTP requests and responses as input and outputs the object-level HTTP information. 
Initially, before receiving any request, the match module will ignore all response inputs and stay at state $-1$. 
After the first request is received, the algorithm comes to state $1$. 
The definition for state $1$ is that the number of requests is greater than the number of responses, %, and the number of requests is no larger than a threshold $N_\text{request} \in \mathbb{Z}^{+}$. 
%Here, $N_\text{request}$ denotes the maximum level of multiplexed HTTP objects that our algorithm can take, which is also the maximum number of individual HTTP objects in each super object. In the example of Fig. \ref{fig_match}, we set $N_\text{request} = 5$, which means our algorithm allows up to five individual request-response pairs to be grouped into one super HTTP object.  
%Staying at state $1$ means
meaning that some HTTP requests have been sent out, but not all of their responses are  received, therefore the algorithm waits at state $1$ for more responses to finish the request-response match. 
Once the match module receives enough responses so that the number of requests becomes less than or equal to the number of responses, it  moves to state $2$. 
However, at state $1$, if no request or response is received within $20$ RTTs, then the match module is timeout, and the algorithm moves to state $0$ to output an HTTP object consisting of all received requests and responses.  

State $2$ means the match module has received at least equal numbers of requests and responses, which is enough for one-to-one request-response match. 
However, at this moment, it is uncertain 
%state $2$, no output will be given, because we are not sure 
whether there will be more response objects coming, due to re-transmission and mis-estimation.  
Thus, the module waits at state $2$ for one RTT. %either  more response, or timeout, or a new request. 
Any new response arrives within one RTT will be added into the current HTTP object. 
If no new response arrives within one RTT, the current module is timeout, and moves to state $0$ to output the estimated HTTP object. 
If any new request is received at state 2, it will be hold until timeout, to form the next HTTP object.  
\begin{comment} 
At state $2$ if a new request comes, the algorithm will move to another waiting state $3$ without an output. 
The information of the new request will be processed separately from the previous requests and responses, and its information be hold into a new HTTP object.  
The purpose of state $3$ is to make sure in the following one RTT, no further response arrives, before we output the current HTTP estimation.  
Therefore, the match module will stay at state $3$ for  one RTT. 
During this time period, if a new response arrives, then the time difference between this response with the new request is less than one RTT, therefore this response must be associated with the previous requests that are received before state $2$. 
Meanwhile, during the one-RTT period at state $3$, if more new requests arrive, these requests will be hold into the new HTTP object, and the algorithm stays at state $3$ until one-RTT timer expires. 
After one RTT past, an estimated HTTP object will be output combining the previous requests received before state $2$ and all received responses, and then the match module will move to state $1$. % is the number of new requests received at states $2$ and $3$ is greater than one, or move to state $2$ is the number of new requests is equal to one. 
Note that, all newly received requests at both state $2$ and state $3$ will be associated with a new HTTP object, and the purpose of state $3$ is to make sure that the time difference between a request and its associated response is at least one RTT.  
\end{comment}
Lastly, state $0$ is the idle state, in which %the QUIC connection has been quite for more than $20$ RTTs. 
%At state $0$, 
all responses will be discarded, and if a request is received, the match module moves to state $1$. 

The match module takes all estimated requests and responses as input, and generates the HTTP request-response objects as output, % An example of  flow chart for the matching algorithm is given in Fig. \ref{fig_matchFlowReq} and Fig. \ref{fig_matchFlowResp}, where  Fig. \ref{fig_matchFlowReq} shows the match approach given a request input, and Fig. \ref{fig_matchFlowResp} is for a response input.  
%In the end, Fig. \ref{fig_overall} clearly presents the relationship between the aforementioned request module, response  module, and  matching module. %, where the request module takes the request QUIC packets as input and generates request objects as output, the response module takes  the response QUIC packets as input and generates response objects as output, and finally, the match module takes the request and response object as input, and produces the object-level HTTP output. 
% In the end, 
while the connection-level output can be calculated, combing all estimated object-level information. 
% based on all object-level output over the considered QUIC connection. 

\begin{table*}
	\centering
	\caption{\label{table_per} Performance summary} \vspace{-0.1cm}
	\renewcommand{\arraystretch}{1.3}  
	\begin{tabular}{|c|c|c|c|c|c|c|c|} 
		\hline
		& \textbf{Dataset}  & \textbf{Match} & \textbf{Request start} & \textbf{Request size} & \textbf{Response start} & \textbf{Response end} & \textbf{Response size}\\
		&   & \textbf{accuracy} & \textbf{time error} & \textbf{accuracy} & \textbf{time error} & \textbf{time error} & \textbf{accuracy}\\
		\hline
		& Comcast, Chrome, small-scale& $96\%$ & 0  & $99\%$ & $\leq 20\%$ RTT &  $50\%$ RTT & $97\%$\\ 
		\textbf{Youtube}	& HughesNet, Chrome, small-scale & $95\%$ &  $\leq 1\%$ RTT  & $98\%$ & $\leq 15\%$ RTT & $\leq 10\%$ RTT & $98\%$\\
		& HughesNet, Firefox, small-scale & $92\%$ & $\leq 10\%$ RTT  & $95\%$ & $\leq 15\%$ RTT & $\leq 10\%$ RTT & $99\%$\\
		& HughesNet, Chrome, large-scale & $94\%$ &  $\leq 15\%$ RTT  & $96\%$ & $\leq 25\%$ RTT & $\leq 20\%$ RTT & $97\%$\\ 
		\hline
		\textbf{Google}	  	& Comcast, Chrome, small-scale & $93\%$ & $ \leq $ one RTT & $97\%$ &  $ \leq 50\%$ RTT & $\leq $ one RTT & $95\%$ \\
		\textbf{drive}		& HughesNet, Chrome, small-scale& $96\%$ &  $\leq 10\%$ RTT & $91\%$ &  $\leq 10\%$ RTT &  $\leq 15\%$ RTT & $99\%$ \\
		\textbf{login}		& HughesNet, Firefox, small-scale& $99\%$ & $\leq 10\%$ RTT	& $99\%$ &  $\leq 10\%$ RTT &  $\leq 15\%$ RTT & $91\%$ \\
		\hline
		\textbf{Google}		& Comcast, Chrome, small-scale& $87\%$ & $\leq 1\%$ RTT & $85\%$ & $\leq 5\%$ RTT  &  $\leq 1\%$ RTT  & $94\%$  \\
		\textbf{drive}		& HughesNet, Chrome, small-scale& $88\%$ & $\leq 50\%$ RTT & $89\%$ &  $\leq 50\%$ RTT  &  $\leq 50\%$ RTT  & $85\%$   \\
		\textbf{download} 	& HughesNet, Firefox, small-scale& $85\%$ & 0              & $99\%$ &  $\leq 10\%$ RTT  &  $ \leq 5\%$ RTT  & $99\%$   \\
		\hline
		\textbf{Google}  	& Comcast, Chrome, small-scale&  $93\%$ &  10 RTTs & $78\%$ & 3 RTTs  &  one RTT & $97\%$ \\ 
		\textbf{drive}		& HughesNet, Chrome, small-scale& $96\%$ &    $\leq 50\%$ RTT & $77\%$ &  $\leq 20\%$ RTT  &  $\leq 30\%$ RTT & $99\%$ \\ 
		\textbf{upload}  	& HughesNet, Firefox, small-scale& $92\%$ &  $\leq 50\%$ RTT & $75\%$  & $\leq 10\%$ RTT  & $\leq 1\%$ RTT & $99\%$ \\ 
		\hline
		\textbf{Facebook}/ 	& Comcast, Chrome, small-scale& $100\%$ & 0  & $100\%$ & $\leq 5\%$ RTT & 0  & $99\%$\\
		\textbf{Instagram}/	& HughesNet, Chrome, small-scale& $100\%$ &  0  & $97\%$ &  $\leq 15\%$ RTT &  $ \leq 20\%$ RTT  & $99\%$\\
		\textbf{Google}     & HughesNet, Firefox, small-scale& $97\%$ & 0  & $99\%$ & $ \leq 20\%$ RTT &  $\leq 10\%$ RTT  & $94\%$\\
		\hline
	\end{tabular}  \vspace{-0.1cm}
	%\end{center}
\end{table*}

\subsection{Supporting modules} \label{sec_algo_support}

To enable the proposed algorithm to work in different networks under various communication conditions,  three supporting modules are introduced to adjust key parameters. % in our algorithm. 
%The supporting modules enable the auto-adjustment of length threshold for QUIC data packets, auto-detection of MTU size, and auto-estimation of RTT, respectively.  
%Note that, for both length threshold and MTU size modules, our estimation will give the real-time values for both upstream and downstream separately over each QUIC connection.  

\subsubsection{Dynamic threshold of data packet length} \label{sec_algo_thre}

For the request data packet, the initial length threshold is $L_{\text{req}} = 50$, i.e., a QUIC packet from the client to server with a length smaller than $50$ bytes will be considered as a non-data packet. 
Generally, it is easy, using $L_{\text{req}}$, to detect  the non-data packet with a fixed or typical length, such as control frame.   
However, for ACK packets with variant sizes, a dynamic threshold is needed. 
Assume the downstream from server to client experiences packet loss, the client will inform the server with the packet missing information in the ACK packet. 
If the number of lost response packets keeps increasing, the ACK packet size from the client to server will become larger. Once ACK length %, as more information needs to be given back to the server.  
% Once the ACK size 
comes to $50$ bytes,  the initial threshold $L_{\text{req}}$ can no longer work. % separate ACK  and data packets anymore. 
%To address the large ACK size in face of packet loss, %we need to monitor the real-time length of the ACK packets. 
%Thus, 
%an auto-adjustment method of request size threshold is designed as follows:  
% Given that 

Since the ACK packet size increases gradually,  we can track the length change and adjust the threshold accordingly. 
%Therefore, At the beginning, we use the initial threshold $L_{\text{req}} = 50$ to separate small-size ACK packets from the large-sized request packets. 
For example, over a QUIC connection, once ten non-data packets have been detected, the middlebox can take the maximum length of the last ten small-sized packets as $l^{max}_{ack} = \max \{l^{1}_{ack}, \cdots, l^{10}_{ack}\} $, and adjust the request threshold by $L_{\text{req}} = l^{max}_{ack} + 10$. % l^{2}_{ack}, %, where a suitable value for $M$ and $\Delta l$ is $M=10$ and $\Delta l = 10$, respectively.   
In the following communication, the maximum length of the latest ten non-data packets $l^{max}_{ack}$ will be updated for every detected non-data packet, and the request threshold is updated accordingly. % as $L_{\text{req}}= l^{max}_{ack} + 10$. 
A similar rule applies to the response packet length, where the initial threshold is $L_{\text{resp}} = 35$; after ten non-data response packets are detected,  the response threshold is updated by the maximum length of the latest ten non-data packets, plus ten bytes.   
Based on our analysis, the proposed  scheme shows almost $100\%$ accuracy to separate the ACK packets from the data packets for both QUIC request and response estimations.  %dynamic threshold

\subsubsection{Auto-detection for MTU size}  \label{sec_algo_support_MTU}

The MTU size of both QUIC and UDP packets depends on the network setting, server implementation, and client device type. 
Therefore,  MTU can take different values for different QUIC connections, or over the same connection but in different communication directions.  
The auto-detection algorithm for MTU size is designed as follows: 
The initial MTU-value for QUIC packets is set to be $L_{\text{MTU}} = 1200$.  
Next, for each packet, the MTU value will be updated by taking the maximum out of the length of the new packet and the current MTU value.   
%A similar update will be applied to the server-to-client MTU for each response packet.   
%Based on our analysis, i
In most cases, the MTU values for both directions over a QUIC connection can be accurately detected within the handshake stage.

\subsubsection{RTT estimation}  \label{sec_algo_support_rtt}

As shown both in Fig. \ref{fig_time} and Fig. \ref{fig_0rtt}, the QUIC handshake stage requires the client to starts with a Client Hello packet, and then, the server will reply with Server Hello. This round-trip pattern during the handshake stage provides a chance for RTT estimation. 
Especially when the QUIC connection is established without previous memory, handshake stage usually involves more than one round-trip, then the value of RTT can be calculated by averaging the time spent over these round-trips during handshake.

\section{Performance Evaluations} \label{sec_performance}

In this section, we evaluate the performance of the proposed algorithm, using the QUIC trace collected from various network environments.
%To evaluate the performance of the QUIC characterization algorithm, we collected one small-scale and one large-scale datasets at the client side, under a variety of experiment settings. % for different purposes. % using  Wireshark as network analyzer. % to collect packet-level information.   
%Note that, both the ground-truth and estimation results of our proposed algorithm are evaluated using the packet level information in this section. 
%First, in the small-scale data collection, we applied two browser softwares, over two network systems, using two operation systems, and collected QUIC trace for five different web application services. 
In particular, we applied Chrome and Firefox as client browsers on both Windows and Linux operation systems, over HughesNet satellite system and Comcast terrestrial system, to collect QUIC traces for video traffic, web-browsing, user login authentication,  file upload, and download traffic. 

In the small-scale collection, we used Wireshark to manually collect QUIC traffic, and decrypted packets by setting the SSLKEYLOGFILE environment variable. %, under each setting.  
For the large-scale collection, we applied Puppeteer as a high-level API to control Chrome and play Youtube videos following some given playlists, and used tcpdump  to collect packet-level QUIC trace. % for Youtube traffic. %as a data-network packet analyzer 5information to get 
The large-scale dataset is limited to web browsing and video traffic from Youtube, over HughesNet satellite system, using Chrome as browser in the client-side Linux operation system. 
We run the large-scale data collection continuously for 11 days from Feb 9 to Feb 20, 2023, resulting in over $1,000$ times of video plays with over $11,000$ TCP connections and $18,000$ QUIC connections between our client browser with over $400$ server IP addresses.

Table \ref{table_per} shows the evaluation performance results over the small-scale Comcaset dataset, small-scale HughesNet dataset over Chrome and Firefox, and large-scale HughesNet dataset for Youtube traffic, respectively. 
 %Note that, in the online estimation, the proposed algorithm receives only one QUIC packet at each time,  while in the offline estimation, the proposed algorithm can see all QUIC packets' information over a QUIC connection. 
%while Table \ref{table_on_J2} and Table \ref{table_on_large_scale} gives the evaluation results for the small-scale and  large-scale satellite data, respectively. % and in the end, Table  shows the  metrics results of online algorithm for youtube  over Jupiter II satellite. 
First, our algorithm yields a high matching accuracy of over $85\%$, for all types of web traffic, in all environment settings.  
In the request estimation, other than the upload traffic, the proposed method shows an accurate estimation result, where the request start time error is 

\cleardoublepage
\newgeometry{left=0.625 in, right=0.645in, top=0.75 in, bottom=1.05 in}
%\printbibliography

\noindent smaller than one RTT, and the request size accuracy is higher than $85\%$. 
Different from other traffic types with small-sized requests and large-sized responses, the file upload shows a reversed pattern, where the traffic from client to server is much more than the data from server to client. This uncommon pattern results in a lower accuracy of $75\%$ in the request size estimation, and up to $10$ RTTs error in the request start time estimation. In our future work, we will further refine the algorithm design, by adding more waiting states in the request state machine, to improve the request estimation for bulk upload traffic. 
Similarly, the response estimation shows a satisfied result of time error small than one RTT,  and size accuracy of over $85\%$, for all web services under all settings, except for bulk upload. 
%Furthermore, for the response end time, most traffic types yield an acceptable estimation error of less than half RTT, except for use login, whose traffic requires more authentication in the HTTP request and response transmission, thus it brings more challenges for our  algorithm to distinguish the response data packet from the control packets with authentication information. Therefore, the estimated response end time for use login can yield an error of up to $10$ RTTs.  
%Meanwhile, the performance of the online algorithm in Talbe \ref{table_on_comcast} shows a similar result as its offline version, where the proposed QUIC characterization algorithm gives highly accurate estimations for most of the web application services, except for the upload traffic. 
%Next, Table \ref{table_on_large_scale} shows that the proposed algorithm has  a satisfied estimation over the QUIC trace collected in the Jupiter II satellite networks, where all the match and size estimations have an accuracy above $94\%$, and  all the time estimation errors are less than $35\%$ RTT. 

Note that, compared with the terrestrial Comcast network, the satellite system has a much larger RTT due to the long propagation distance between the ground terminal/gateway and the geostationary satellite. %Therefore, when we count the percentage-wise time error of a consider network, the evaluation results over HughesNet system usually yield a smaller relative error given its larger RTT, compared with Comcast system.  
Therefore, the evaluation results prove that  our proposed algorithm can work in various networks while guaranteeing an accurate estimation result. 
Furthermore, due to the limited space, Table \ref{table_per} only shows six key performance metrics from Table \ref{tab_output}, for online estimation only. 
Note that, the offline algorithm yields a similar estimation accuracy, and the other performance metrics also show satisfied results.

\section{Limitation and Future Work}\label{discusison}

Given the empirical nature of the proposed algorithm, one limitation of our work is the performance degradation in face of excess packet loss. Massive packet loss yields a lot of data re-transmission, so that the typical transmission pattern cannot be recognized; also, the ACK packets with large sizes will be confused with the data packets, even with a dynamic length threshold. 
Meanwhile, the proposed algorithm only provides a coarse-grained estimation for interleaved HTTP request-response objects, since as an ISP with limited information visible in the encrypted QUIC packets, it is impossible to distinguish individual request-response pairs using interleaved timeline with length and order information only. Thus, grouping the multiplexed objects into a super HTTP object is the best estimation we could make. 
Furthermore, if client or server implementations apply padding as a countermeasure of traffic analysis, then the length of all QUIC packets will be MTU. In  this case, our proposed algorithm might fail, given only time and order information available. 

In the future work, we will apply the estimated object-level and connection-level HTTP information for network operation and management, including the traffic classification and  QoE estimation. 
For example, different web traffics have distinct HTTP patterns, where a video connection requests content data periodically resulting in a clear and separable request-response  pattern as shown in Fig. \ref{fig_time}, while a web-browsing connection requests different types of content at the same time, inducing interleaved requests and responses.   %including text, image, HTML, CSS and Javascript
%As a consequence, the web-browsing QUIC connection usually shows a higher  level of HTTP multiplexing than a video connection. 
Such pattern difference enables the ISPs to classify each QUIC connection into different application categories.  
%In our previous work \cite{jain2019application}, a rule-based traffic classification algorithm has been proposed. 
%However, this work is only applicable to TCP protocol, but not QUIC. 
%In order to incorporate the full-encryption feature of QUIC, we can rely on the object-level, connection-level, as well as application-level output of this work to design algorithms for QUIC traffic classifications.  
Moreover, the application-layer information can be applied to infer the user's QoE over the encrypted QUIC connection. 
For example, the download rate per object can be calculated by response size over response duration, and the time-to-first-byte can be evaluated via the estimated request start time and the response start time.

\section{Conclusion} \label{sec_conclusion}

In this work, we have analyzed the characteristics of QUIC traffic, by  passively monitoring the QUIC encrypted packets to infer the application-layer attributes.  
To this end, we have studied the rationale of QUIC protocol design, and summarized the key pattern for HTTP request and response communications over QUIC protocol. 
By carefully choosing the time and size features which are still visible in the encrypted QUIC packets, we have designed a novel rule-based algorithm to estimate the attributes of HTTP requests and responses. %  from the HTTP object level, QUIC connection level, and application-layer service level. 
%The proposed algorithm consists of three modules to address the request estimation, response estimation, and request-response match, respectively. 
The performance evaluation showed satisfactory results in different network systems for various web applications. 
%This work can be further extended to support  traffic classification and QoE estimation for ISPs to manage the encrypted QUIC traffic in their network systems.  

%\def\baselinestretch{0.95}

\bibliographystyle{IEEEtran}
\bibliography{bibliography}

\begin{thebibliography}{10}
\providecommand{\url}[1]{#1}
\csname url@samestyle\endcsname
\providecommand{\newblock}{\relax}
\providecommand{\bibinfo}[2]{#2}
\providecommand{\BIBentrySTDinterwordspacing}{\spaceskip=0pt\relax}
\providecommand{\BIBentryALTinterwordstretchfactor}{4}
\providecommand{\BIBentryALTinterwordspacing}{\spaceskip=\fontdimen2\font plus
\BIBentryALTinterwordstretchfactor\fontdimen3\font minus
  \fontdimen4\font\relax}
\providecommand{\BIBforeignlanguage}[2]{{%
\expandafter\ifx\csname l@#1\endcsname\relax
\typeout{** WARNING: IEEEtran.bst: No hyphenation pattern has been}%
\typeout{** loaded for the language `#1'. Using the pattern for}%
\typeout{** the default language instead.}%
\else
\language=\csname l@#1\endcsname
\fi
#2}}
\providecommand{\BIBdecl}{\relax}
\BIBdecl

\bibitem{akbari2021look}
I.~Akbari, M.~A. Salahuddin, L.~Ven, N.~Limam, R.~Boutaba, B.~Mathieu,
  S.~Moteau, and S.~Tuffin, ``A look behind the curtain: traffic classification
  in an increasingly encrypted web,'' in \emph{Proceedings of the ACM on
  Measurement and Analysis of Computing Systems}, vol.~5, no.~1.\hskip 1em plus
  0.5em minus 0.4em\relax ACM New York, NY, USA, 2021, pp. 1--26.

\bibitem{langley2017quic}
A.~Langley, A.~Riddoch, A.~Wilk, A.~Vicente, C.~Krasic, D.~Zhang, F.~Yang,
  F.~Kouranov, I.~Swett, J.~Iyengar \emph{et~al.}, ``The {QUIC} transport
  protocol: Design and {I}nternet-scale deployment,'' in \emph{Proceedings of
  the conference of the ACM special interest group on data communication},
  2017, pp. 183--196.

\bibitem{xu2020csi}
S.~Xu, S.~Sen, and Z.~M. Mao, ``{CSI}: Inferring mobile {ABR} video adaptation
  behavior under {HTTPS} and {QUIC},'' in \emph{Proceedings of the Fifteenth
  European Conference on Computer Systems}, 2020, pp. 1--16.

\bibitem{quicUsage}
\BIBentryALTinterwordspacing
``Usage statistics of {QUIC} for websites,'' accessed: May, 2023. [Online].
  Available: \url{https://w3techs.com/technologies/details/ce-quic}
\BIBentrySTDinterwordspacing

\bibitem{border2020evaluating}
J.~Border, B.~Shah, C.-J. Su, and R.~Torres, ``Evaluating {QUIC}’s
  performance against performance enhancing proxy over satellite link,'' in
  \emph{Proceedings of the IEEE IFIP Networking Conference}, 2020, pp.
  755--760.

\bibitem{anderson2019limitless}
B.~Anderson, A.~Chi, S.~Dunlop, and D.~McGrew, ``Limitless {HTTP} in an {HTTPS}
  world: Inferring the semantics of the {HTTPS} protocol without decryption,''
  in \emph{Proceedings of the Ninth ACM Conference on Data and Application
  Security and Privacy}, 2019, pp. 267--278.

\bibitem{jain2019application}
K.~Jain and C.-J. Su, ``Application characterization using transport protocol
  analysis,'' Oct.~22 2019, {US} Patent 10,454,804.

\bibitem{zhan2021website}
P.~Zhan, L.~Wang, and Y.~Tang, ``Website fingerprinting on early {QUIC}
  traffic,'' \emph{Computer Networks}, vol. 200, p. 108538, 2021.

\bibitem{tong2018novel}
V.~Tong, H.~A. Tran, S.~Souihi, and A.~Mellouk, ``A novel {QUIC} traffic
  classifier based on convolutional neural networks,'' in \emph{Proceedings of
  the IEEE Global Communications Conference (GLOBECOM)}, 2018, pp. 1--6.

\bibitem{mangla2019using}
T.~Mangla, E.~Halepovic, M.~Ammar, and E.~Zegura, ``Using session modeling to
  estimate {HTTP}-based video {QoE} metrics from encrypted network traffic,''
  \emph{IEEE Transactions on Network and Service Management}, vol.~16, no.~3,
  pp. 1086--1099, 2019.

\bibitem{mazhar2018real}
M.~H. Mazhar and Z.~Shafiq, ``Real-time video quality of experience monitoring
  for {HTTPS} and {QUIC},'' in \emph{Proceedings of the IEEE Conference on
  Computer Communications (INFOCOM)}, 2018, pp. 1331--1339.

\bibitem{bentaleb2020inferring}
A.~Bentaleb, S.~Harous \emph{et~al.}, ``Inferring quality of experience for
  adaptive video streaming over {HTTPS} and {QUIC},'' in \emph{Proceedings of
  the IEEE International Wireless Communications and Mobile Computing (IWCMC)},
  2020, pp. 81--87.

\bibitem{kuehlewind2020manageability}
M.~Kuehlewind and B.~Trammell, ``Manageability of the {QUIC} transport
  protocol,'' \emph{Internet Engineering Task Force, Internet-Draft
  draft-ietfquic-manageability-06}, 2020.

\bibitem{iyengar2020quic}
J.~Iyengar, M.~Thomson \emph{et~al.}, ``{QUIC}: A {UDP}-based multiplexed and
  secure transport,'' \emph{Internet Engineering Task Force, Internet-Draft
  draft-ietf-quic-transport-27}, 2020.

\end{thebibliography}

\begin{comment}

\begin{figure*} \vspace{-0cm}
	\center 
	\includegraphics[width=12.5cm]{overall.png}
	\caption{\label{fig_overall}  The overall algorithm. }
\end{figure*}

\begin{figure*}
	\center 
	\includegraphics[width=12.8cm]{request_flow.png}
	\caption{\label{fig_reqFlow}  The flow chart for the request estimation algorithm. }
\end{figure*}

\begin{figure*}
	\center 
	\includegraphics[width=13.5cm]{response_flow.png} \vspace{0.1cm}
	\caption{\label{fig_respFlow}  The flow chart for the response estimation algorithm. }
\end{figure*}

\begin{figure*}
	\center 
	\includegraphics[width=12 cm]{match_request_flow.png}  
	\caption{\label{fig_matchFlowReq}  The flow chart for the match model given an input of a request. }
\end{figure*}

\begin{figure*}
	\center 
	\includegraphics[width=13cm]{match_response_flow.png}  
	\caption{\label{fig_matchFlowResp}  The flow chart for the match model given an input of a response. }
\end{figure*}
\end{comment}

\end{document}